\documentclass[twocolumn,prb,aps,ams,epsfig,graphicx,showpacs,floatfix]{revtex4}
\usepackage{graphicx}
\newcommand{\be}{\begin{equation}}
\newcommand{\ee}{\end{equation}}

\newcommand{\bea}{\begin{eqnarray}}
\newcommand{\eea}{\end{eqnarray}}
\newcommand{\bd}{\begin{displaymath}}
\newcommand{\ed}{\end{displaymath}}
\newcommand{\bi}{\begin{itemize}}
\newcommand{\ei}{\end{itemize}}
\newcommand{\bc}{\begin{center}}
\newcommand{\ec}{\end{center}}
\newcommand{\bfl}{\begin{flushleft}}
\newcommand{\efl}{\end{flushleft}}
\newcommand{\bfr}{\begin{flushright}}
\newcommand{\efr}{\end{flushright}}
\newcommand{\f}{\frac}

\def\br{{\bf r}}\def\bR{{\bf R}}
\def\bk{{\bf k}} \def\bK{{\bf K}} \def\bp{{\bf p}}

\def\6{\partial}  \def\b{\beta}
 \def\d{\delta} \def\ve{\varepsilon}

\def\ss{\sigma} 

  \def\D{\Delta}

\def\={\!\!\!&=&\!\!\!}
\def\+{\!\!\!&&\!\!\!+~}
\def\-{\!\!\!&&\!\!\!-~}

\begin{document}

\author{Ionel \c{T}ifrea\cite{adresa} and Michael E. Flatt\'{e}}
\affiliation{Department of Physics and Astronomy and Optical
Science and Technology Center, University of Iowa, Iowa City
52242, USA}
\date{\today}
\title{Dynamical nuclear polarization and nuclear
magnetic fields in semiconductor nanostructures}

\begin{abstract}
We investigate the dynamic nuclear polarization from the hyperfine
interaction between nonequilibrium electronic spins and nuclear
spins coupled to them in semiconductor nanostructures. We derive
the time and position dependence of the induced nuclear spin
polarization and dipolar magnetic fields. In GaAs/AlGaAs parabolic
quantum wells the nuclear spin polarization can be as high as 80\%
and the induced nuclear magnetic fields can approach a few gauss
with an associated nuclear resonance shift of the order of kHz
when the electronic system is 100\% spin polarized. These fields
and shifts can be tuned using small electric fields. We discuss
the implications of such control for optical nuclear magnetic
resonance experiments in low-dimensional semiconductor
nanostructures.
\end{abstract}
%
\maketitle

\section{Introduction}

Coherent control of the spin degrees of freedom in low dimensional
semiconductor structures may lead to spin based electronic devices
and quantum information processing \cite{book,wolf}. The practical
realization of quantum computing requires the preparation,
manipulation, and measurement of pure quantum states
\cite{Loss-DiVincenzo,kane,taylor}. Nuclear spins are ideal
candidates, as all required conditions can be achieved based on
the hyperfine interaction between electronic and nuclear spins
\cite{taylor,gershenfeld}. Despite the local character of the
hyperfine interaction, single nuclear spin manipulation is hard to
achieve, an inconvenience which can be overcome by using instead
nuclear spin clusters \cite{meier}. For GaAs quantum wells and
quantum dots the nuclear spin coherence time can be as long as a
second \cite{barrett,tifrea}, much longer than the electron spin
coherence time, of the order of 100 ns
(Ref.~\onlinecite{kikkawa}).

Control over the nuclear spin dynamics in semiconductor
nanostructures is realized by various methods. Control of
collective excitations\cite{smet,desrat} can modify the enhanced
nuclear spin relaxation times in a GaAs quantum well (QW).
Adjacent ferromagnetic layers can ``imprint" nuclear
spin\cite{kawakami} in $n$-type GaAs QW's. A flexible method of
nuclear spin manipulation, using gate voltages to electrically
address a wide distribution of polarized nuclei within an AlGaAs
parabolic quantum well (PQW), was recently
demonstrated\cite{martino}. Optically injected spin polarized
electrons transfer their spin polarization to the nuclear
population via dynamic nuclear polarization (DNP)
\cite{overhauser}, resulting in a position dependent nuclear
polarization within the PQW. Gate voltages are then used to shift
the electron population and thus produce polarized nuclei with
different probabilities at various positions in the PQW. The
position dependent nuclear polarization was measured by time
resolved Faraday rotation (TRFR) experiments, which
showed\cite{martino} that a 8 nm wide distribution of polarized
nuclei can be manipulated electrically over a range of 20 nm.

Here we derive general formulas describing the nuclear
polarization, and resulting nuclear dipolar fields, achieved
dynamically in low dimensional semiconductor nanostructures due to
the hyperfine interaction between electronic and nuclear spins.
Just as was found for nuclear and electron relaxation
times\cite{tifrea}, the central physical quantity determining the
nuclear polarization is the electronic local density of states
(ELDOS) at the nuclear position. The position dependence of the
induced nuclear polarization in semiconductor nanostructures is
shown to be a function of the initial polarization of the
electronic population and various nuclear interactions which lead
to nuclear spin relaxation. We calculate how the nuclear
polarization within the semiconductor nanostructure can be
manipulated with electric fields by changing the ELDOS at
particular locations.  Our results are relevant for nuclear
magnetic resonance (NMR) and TRFR experiments in semiconductor
nanostructures \cite{marohn,harley,salis}. For AlGaAs PQW's we
propose an experimental setup where the efficiency of optical DNP
will be enhanced by the proper insertion of a $\delta$-doped layer
of different nuclei at a specific position. Calculations of the
induced nuclear polarization also allow us to predict the nuclear
magnetic resonance shift in semiconductor nanostructures.

The paper is organized as follows. In the next Section we present
a theoretical derivation of DNP in samples with reduced
dimensionality. Section III presents numerical estimations for the
induced nuclear spin polarization and the resulting dipolar
nuclear magnetic fields for an AlGaAs PQW. Section IV
gives our conclusions.

\section{Dynamical Nuclear Polarization}

Dynamical nuclear polarization was theoretically described by
Overhauser \cite{overhauser} in bulk metallic samples. The
interaction between nuclear and electronic spins leads to an
enhanced nuclear spin polarization which can be measured in NMR
and TRFR experiments. For semiconductor bulk materials such as
GaAs, the DNP effect can be enhanced via optical
techniques\cite{paget,kikkawasci}. The same optical pumping
technique was successfully used to polarized nuclei in quasi
two-dimensional semiconductor heterostructures\cite{berg,barrett}.
In this Section we investigate theoretically DNP in samples with
reduced dimensionality such as quantum wells and quantum dots
(QD).

The hyperfine interaction between electronic and nuclear spins is
described by the Hamiltonian
\be\label{intham}
H_n=\f{8\pi}{3}\;\b_e\b_n\left({\vec \ss_n }\cdot{\vec
\ss_e}\right)\;\d(\br-\br_n)\;,
\ee
where $n$ labels the nuclei, $\b_n$ and $\b_e$ are the nuclear and
electron magnetic moments, ${\vec \ss_n}$ and ${\vec \ss_e}$ are
the Pauli spin operators for the nucleus and electron, $\br-\br_n$
represents the relative distance between the nuclear and
electronic spins, and $\d(\br)$ is the Dirac delta function. The
Hamiltonian describes a flip-flop process for both electronic and
nuclear spins in which the energy and the total angular momentum
are conserved. We consider the interaction to be weak, so we can
use perturbation theory to describe its effects. To understand the
dynamics of the electronic and nuclear spins governed by the
hyperfine interaction we consider the system to be in an external
magnetic field, $B_0$, which partially orients the spins. We
assume dephasing of electronic orbital information on timescales
much faster ($\sim 100$fs) than either the precession of electron
spins in momentum-dependent effective magnetic
fields\cite{Dresselhaus} or nuclear decoherence times; this
permits us to neglect the momentum-dependent spin splitting of
electronic states from the spin-orbit
interaction\cite{Dresselhaus}.  The electronic spin polarization
is assumed the same everywhere, described by the spin-up and
spin-down populations $N_+$ (parallel to the applied field) and
$N_-$ (antiparallel to the applied field), respectively. On the
other hand, the nuclear system will develop a position dependent
polarization described by $M_m(\br_n)$, $m=I,I-1,\dots, -I$, where
$I$ is the nuclear spin quantum number. The hyperfine interaction
will act to relax both the electronic and the nuclear spins,
according to the following two equations \cite{tifrea2}
\be\label{Dfinal}
\f{dD}{dt}=\f{D_0-D}{T_{1e}}+
\sum_n\f{\D_0(\br_n)-\D(\br_n)}{T_{1n}(\br)}\;
\ee
and
\be\label{cons}
\f{dD}{dt} = -\f{2I(I+1)(2I+1)}{3}\sum_n \f{d\D(\br_n)}{dt}\;.
\ee
Here $D=N_+-N_-$, $\D(\br_n)=M_{m+1}(\br_n)-M_m(\br_n)$, and $D_0$
and $\D_0(\br_n)$ are their thermal equilibrium values. The
electronic and nuclear spin relaxation times are given by (see
Refs. \onlinecite{tifrea,tifrea2})
\be\label{teQW}
T^{-1}_{1e}=\f{1}{V}\sum_{n}\f{1024\pi^3\b_e^2\b_n^2\int d\ve
A_e^2(\br_n,\ve) f_{FD}'(\ve)}{9\hbar (2I+1)\int d\br d\ve
A_e(\br,\ve) f_{FD}'(\ve)}
\ee
and
\be\label{tnQW}
T^{-1}_{1n}(\br_n)=\f{512\pi^3\b^2_e\b^2_nk_BT\int d\ve
A_e^2(\br_n,\ve) f_{FD}'(\ve)}{3\hbar I(I+1)(2I+1)}\;,
\ee
where $A_e(\br_n,\ve)$ represents the ELDOS, $T$ is the
temperature, and $f_{FD}(\ve)$ the Fermi-Dirac distribution
function. The ELDOS at the nuclear position $\br_n$ is
\be
A_e(\br_n,\ve)=\sum_l |\psi_l(\br_n)|^2\d(\ve-E_l)\;,
\ee
where $l$ labels the state, $E_l$ its energy, and $\psi_l(\br_n)$
its wavefunction at the $n$'th nucleus. Equations (\ref{Dfinal})
and (\ref{cons}) can be combined to obtain a general equation for
the nuclear spin dynamics
\be\label{nuclear}
\f{d\D(\br_n)}{dt}=\f{\D_0(\br_n)-\D(\br_n)}{T_{1n}(\br_n)}
+\f{1}{\left(2I+1\right)k_BT\tilde{N}}\f{D_0-D}{T_{1n}(\br_n)}\;,
\ee
where $\tilde{N}=\int d\br d\ve A_e(\br,\ve) f_{FD}'(\ve)$. The
above equation describes the nuclear spin dynamics due to the
hyperfine interaction. Additionally, the nuclei will relax through
other mechanisms as a result of interactions with phonons,
impurities, electrons, and other nuclei. Such interactions should
be included in any equation for the nuclear spin dynamics, and
they can be included by replacing $1/T_{1n}(\br_n)$ with
$1/T_{1n}(\br_n)+1/T'_n$  in the first term on the right hand side
(rhs) of Eq (\ref{nuclear}). Here $T'_n$ represents the nuclear
spin relaxation time due to additional relaxation mechanisms. Note
that such a replacement is not appropriate for the second term in
the rhs of Eq. (\ref{nuclear}), as this term originates from the
hyperfine interaction alone.

Equation~(\ref{nuclear}) also assumes that nuclear spin diffusion
can be neglected. The nature of the sample determines whether or
not nuclear spin diffusion can be neglected. Paget \cite{paget2}
showed that in bulk GaAs diffusion is very important, and leads to
an uniform polarization of the nuclei across the sample. To
describe diffusion the equation for the time and position
dependence of the nuclear spin polarization has to be modified by
adding a diffusive term. However, often nuclear spin diffusion
appears negligible for low dimensional samples such as QW and
QD\cite{martino,strand}. In the following we will discuss the
consequences of the DNP effect in the absence of nuclear spin
diffusion. Such an assumption should work well for PQW's, the
system for which we will report specific
results\cite{tifrea,martino}.

In DNP, spin polarized electrons created by absorbtion of
polarized light or electrical injection \cite{strand} will
transfer their polarization to the nuclei via the hyperfine
interaction. We assume that the electronic polarization, $D$, is
kept constant by continual resupply of spin polarized electrons.
This would naturally be the case for DC electrical spin injection.
For pulsed optical pumping, however, the spin-polarized electron
population will vary on timescales corresponding to the time
between pulses ($\sim 13$~ns).  Here we rely on the vastly greater
timescales of the nuclei --- as the response times of the nuclei
$(T_{1n})$ are orders of magnitude greater than $13$~ns, the
nuclei see an effective constant average electron spin
polarization. Under these conditions the last term in Eq.
(\ref{nuclear}) is independent of time and the
 time-dependent nuclear polarization is
%
%
\be\label{nuclearpol}
\D(\br_n,t)=
\D_0 + \D_{ind}(\br_n)
\left\{1-\exp{\left[-t\left(\f{1}{T_{1n}(\br_n)}
+\f{1}{T'_n}\right)\right]}\right\}\;,\;\;\;\;\;
\ee
%
where
\be\label{dind}
\D_{ind}(\br_n)=\f{1}{\left(2I+1\right)k_BT\tilde{N}}
\f{T'_n}{T_{1n}(\br_n)+T'_n}(D_0-D)
\ee
represents the induced nuclear polarization due to the hyperfine
interaction. In general the nuclear polarization due to external
magnetic fields, $\D_0$, is about 1\%, suggesting that the large
nuclear polarization originates from the hyperfine interaction.

Two different time regimes can be identified in Eq.
(\ref{nuclearpol}). First, in the initial stages of the DNP
process ($t\ll T_{eff}$; with $T_{eff}^{-1}=T_{1n}^{-1}+T_n^{'-1}$), the
nonequilibrium nuclear system magnetization can be approximated as
\be\label{nuclearshorttime}
\D(\br_n,t)\approx \D_{ind}(\br_n)\f{t}{T_{eff}}\;.
\ee
In general, at low temperatures where the DNP process is
efficient, the relaxation mechanism due to the hyperfine
interaction is the dominant one, making $T_{1n}$ shorter than
$T'_n$. Accordingly, in the initial stage of the DNP process,
$\D(\br_n,t)\propto |\psi_l(\br_n)|^4 t$ (Ref.~\onlinecite{tifrea}). On the
other hand, in the second regime of the DNP process for $t\gg
T_{eff}$, the induced nuclear spin
polarization from the hyperfine interaction will saturate at
\be\label{nuclearsaturated}
\D(\br_n,t)=\D_0+\f{1}{\left(2I+1\right)k_BT\tilde{N}}
\f{T'_n}{T_{1n}(\br_n)+T'_n}(D_0-D)\;.
\ee

\section{Nuclear spin polarization and dipolar magnetic fields}

Large non-equilibrium nuclear polarizations produce real magnetic
fields that act both on the nuclear and on the electronic spins. A
significant effect of these fields is the shift of resonant
frequencies in magnetic resonance experiments associated with
nuclei or electrons. Below we address primarily the effects of DNP
on NMR experiments; their effect on TRFR experiments will be
reported elsewhere \cite{tifrea3}.

We determine the nuclear magnetic fields from the non-equilibrium
occupation of the different states of the nuclear spin due to the
DNP process, and we neglect the equilibrium polarization $\D_0$
from the static magnetic field.  The induced nuclear spin
polarization,
\be
{\cal P}=\f{\sum_m m M_m}{I\sum_m M_m}\;.
\ee
For nanostructured materials ${\cal P}$ will depend on position,
as nuclei in different regions of the sample overlap differently
with the electronic wavefunctions. Also, the time-dependence of
the DNP process will cause  the nuclear spin polarization to
depend on time as well. The induced nuclear magnetization is
\be\label{nuclearmag}
{\cal M}_{ind}(\br_n)=\sum_m m M_m(\br_n)\;.
\ee
The observable physical quantity in NMR experiments, however, is
the nuclear magnetic field produced by this nonequilibrium
magnetization. The position-dependent induced nuclear magnetic
field can be calculated for {\it layered} structures by dividing
the structure into thin slabs stacked in the growth direction and
labeled by $z_n$, and assuming that the nuclei in each slab have
uniform magnetization. Nuclei in different slabs can have
different magnetization due to the potential  dependence of the
nuclear relaxation time on the growth direction\cite{tifrea} due
to a non-uniform ELDOS. The dipolar field from the nuclei, if they
are polarized perpendicular to the growth direction, is
\be\label{dipolarfield}
B_{ind}(\br_n)=\mu_0\mu_n {\cal M}_{ind}(\br_n)\;,
\ee where $\mu_0$ represents the vacuum permeability and $\mu_n$
the nuclear magneton. This magnetic field will act both on the
nuclei and the electrons, and for nuclei the effect can be
measured as a shift in the resonance frequency in an NMR
experiment. This effect is similar to the Knight shift
\cite{slichter}, and can be characterized by
\be\label{shift}
\D\nu(\br_n)=g_n\mu_n B_{ind}(\br_n)\;,
\ee
where $g_n$ is the nuclear g-factor. For low dimensional
nanostructures the shift will depend on position.  The total
nuclear magnetic moment of the sample is
\be\label{average}
M=\int d\br_n \mu_n{\cal M}_{ind}(\br_n)\;.
\ee

For a PQW structure the system's dispersion relations are
quasi-two dimensional, and the total electronic wavefunction is a
product between an envelope function, $\phi(z)$, and a Bloch
function, $u(\br)$,
\be\label{elfunction}
\psi_{j\bK}(\br_n)=\exp{[i\bK\cdot\bR]\;\phi_j(z)\;u(\br_n)}\;.
\ee
Based on this assumption the ELDOS is
\be\label{QWELDOS}
A_e(\br_n,\ve)=\sum_{j} |\phi_{j}({\bf z}_n)|^2\;N_{2D}\Theta(\ve
- E_{j(\bK=0)})\;,
\ee
where $N_{2D}$ is the density of states for a two-dimensional
electron gas and $\Theta(z)$ is the Heavyside step function. In
the following we will consider an Al$_x$Ga$_{1-x}$As PQW ($L=1000$
\AA) with x=0.07 in the center of the structure confined within
two 100 \AA \ Al$_{0.4}$Ga$_{0.6}$As barriers. For this structure
the value of the Bloch function at Ga nuclei was already extracted
in Ref.~\onlinecite{tifrea} and the envelope functions will be evaluated
using a 14-band $\bk\cdot\bp$ calculation \cite{olesberg}. For all
calculations we consider only the first electronic conduction
subband occupied and the electron spin polarization, $D=100\%$,
much greater than the thermodynamic equilibrium one, $D_0$. The
additional nuclear spin relaxation time, $T_n'=600$~s, and is
considered to be temperature independent \cite{mcneil}.

\begin{figure}[t]
\centering
\scalebox{0.8}[0.8]{\includegraphics*{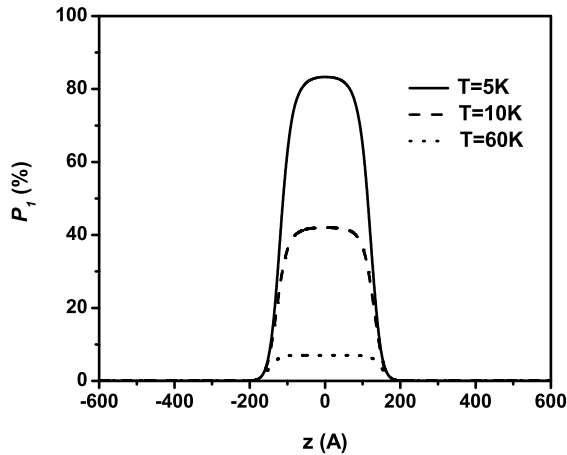}}
%
\caption{The growth-direction position dependence of the saturated
induced nuclear spin polarization in the AlGaAs PQW for $T'_n =
600$ s and different temperatures (full line--$T=5$K, dash
line--$T=10$K, and dash-dot line--$T=60$K).}
\label{fig0}
\end{figure}

Figure~\ref{fig0} presents a quantitative plot of the saturated
induced nuclear spin polarization, ${\cal P}_{sat}(\br_n)$ as a
function of position in the AlGaAs PQW at different temperatures.
For fully polarized electrons ($D=100\%$) the induced nuclear spin
polarization in the center of the PQW can be as high as 80\%,
decreasing drastically on the sides of the sample. The nuclear
spin polarization also decreases as the temperature increases,
making the DNP process effective only at low temperatures. If
higher electronic subbands were considered, the position
dependence of the saturated nuclear spin polarization would change
accordingly.

\begin{figure}[t]
\centering
\scalebox{0.8}[0.8]{\includegraphics*{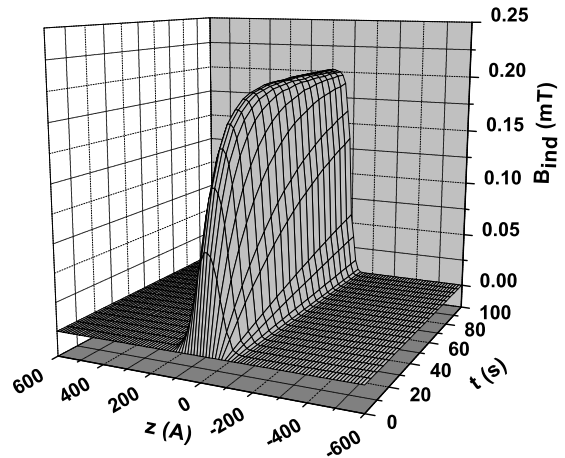}}
%
\caption{The time and position dependence of the induced nuclear
dipolar magnetic field in the AlGaAs PQW for $T'_n = 600$ s.}%
\label{fig1}
\end{figure}

Figure~\ref{fig1} presents a quantitative plot of the induced
nuclear magnetic field $B_{ind}(\br_n,t)$ as a function of time and
position across the AlGaAs PQW for the same situation as in Fig.
\ref{fig0}. The strong confinement, even of this shallow PQW
structure, is reflected in the large induced nuclear magnetic
field at the center. The total response of the sample in NMR experiments
will be mainly due to the central nuclei of the sample, suggesting
that a more effective DNP can be realized by the insertion of
active NMR nuclei at a particular growth-direction position in the sample. For
higher conduction subband occupancy the profile of the induced
nuclear field will change accordingly, as a result of a different
position dependent nuclear spin relaxation time due to the
hyperfine interaction \cite{tifrea}.

\begin{figure}[t]
\centering
\scalebox{0.75}[0.75]{\includegraphics*{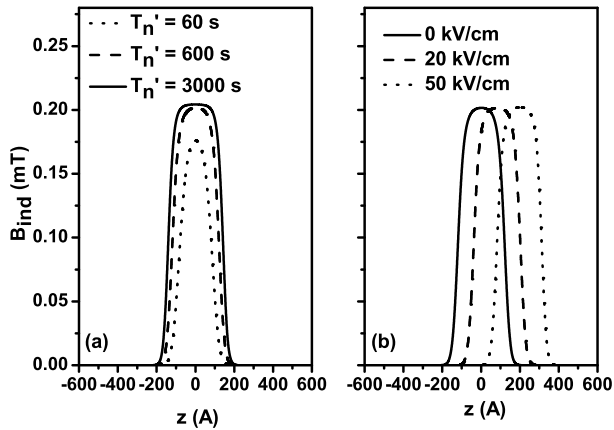}}
%
%
\caption{(a) The saturation value of the induced nuclear
polarization in the AlGaAs PQW for different values of the
additional nuclear relaxation time (full line: $T'_{n}=3000$ s,
dashed line: $T'_{n}=600$ s, and dotted line: $T'_{n}=60$ s). (b)
The saturation value of the induced nuclear polarization in the
AlGaAs PQW for $T'_{n}=600$ s at different values of the applied
electric field (full line: $F = 0$ kV/cm, dashed line: $F = 20$
kV/cm, and dotted line: $F = 50$ kV/cm).} \label{fig2}
\end{figure}

In Fig. \ref{fig2} we present the position dependence of the
saturated induced nuclear magnetic field for different values of
the additional nuclear spin relaxation time, $T'_n$, considering
only the first conduction subband occupied. As seen in Fig.
\ref{fig2}(a) the full width at half maximum (FWHM) of the induced
nuclear magnetic field is strongly dependent on the additional
relaxation mechanisms involving nuclear spins. At low
temperatures, where most of our calculations are performed, the
dominant nuclear spin relaxation mechanism is the hyperfine
interaction, and the measurement of additional nuclear spin
relaxation times is very difficult. Fig \ref{fig2} (b) presents
the saturated induced nuclear magnetic field for $T'_n=600$~s in
the presence of and in the absence of an applied electric field
along the growth direction for the PQW. The control is based on
the manipulation of the ELDOS. Our calculation suggests the
possibility of further controlling and manipulating the nuclear
spin distribution in AlGaAs PQW. When a $\d$-doped layer of active
nuclei is inserted within the PQW, the initial nuclear
polarization of the nuclei in the layer can be directly controlled
by electric fields. A different way to electrically control the
induced position dependent nuclear field would be to gate the PQW
and control the hyperfine nuclear spin relaxation time through the
electronic subband occupancy \cite{tifrea,tifrea2}. Different
shapes and different position dependences of the induced nuclear
field are expected in this case.

\begin{figure}[t]
\centering
\scalebox{0.75}[0.75]{\includegraphics*{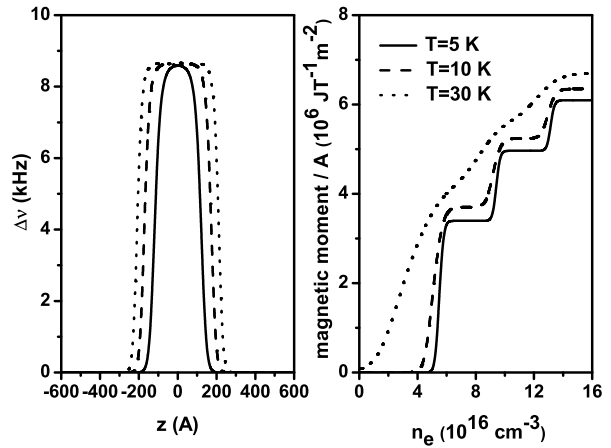}}
\caption{(a) The nuclear resonance frequency shift as function of
position in the PQW for different conduction subband occupancy at
$T$=5 K (full line: single subband occupancy, dashed line: double
subband occupancy, and dotted line: triple subband occupancy). (b)
Total nuclear induced magnetic moment as function of the electron
density for different temperatures (full line: $T$=5 K, dashed
line: $T$=10 K, and dotted line: $T$=30 K).} \label{fig3}
\end{figure}

Fig \ref{fig3}(a) presents the calculated nuclear resonance shift
for the AlGaAs PQW for different conduction subband occupancy at
$T=5$ K. This situation is relevant for a $\d$-doped layer of
active nuclei, and the nuclear resonance shift can reach 8.5 kHz.
Moreover, for a $\d$-doped layer, the resonance shift is fully
controllable with electric fields, both when the field is used as
a control over the electron confinement in the PQW or when the
field is used as a source of subband occupancy in the PQW. Fig
\ref{fig3}(b) shows the total nuclear magnetic moment as a
function of the electron density for different temperatures. As
the electron density increases, the number of occupied conduction
subbands will increase, and accordingly the nuclear magnetic
moment of the well will increase quasi-stepwise. For the
considered PQW the energy difference between the minimum of two
consecutive conduction subbands is about $\D E=15$ meV, meaning
that at $T=30$ K (dotted line) thermal smearing of the Fermi
function will suppress the stepwise shape of the total nuclear
magnetic moment. For PQW's with a greater difference $\D E$ the
stepwise shape will persist even at higher temperatures. The total
nuclear magnetic moment for a fixed electronic density depends on
temperature, as $T_{1n}$ and $T_n'$ have different temperature
dependencies (in Fig. \ref{fig3} (b) we considered $T_{1n}\sim T$
and $T_n'\sim const.$ \cite{tifrea,mcneil}).

\section{Conclusions}

The DNP process is of considerable interest for samples with reduced
dimensionality as it represents a path for highly efficient NMR
and TRFR measurements.  As a result of DNP, nuclei will
produce effective magnetic fields which in turn will act both on the
nuclear and electronic populations. The effects of those magnetic
fields should be observable in NMR and TRFR experiments as shifts
in the resonant frequencies. Usually, as a result of the hyperfine
interaction there will be at least two types of induced nuclear
magnetic fields, a hyperfine nuclear magnetic field and a dipolar
magnetic field. The hyperfine magnetic field acting on the nuclear
population is an effective magnetic field induced by the polarized
electronic population. There will be also a hyperfine field
created by the polarized nuclei acting on the electrons. Such
hyperfine fields will induce a Knight shift in the nuclear
resonant frequencies, and an Overhauser shift in the electronic
resonant frequencies, respectively. On the other hand, the dipolar
magnetic field is a real magnetic field created as a result of
nuclear spin polarization. The dipolar nuclear magnetic field will
be responsible for an additional shift in the resonant frequencies
of both nuclear and electronic systems similar to Knight and
Overhauser shifts, respectively.

For general low dimensional systems we described the dynamics of
the nuclear spins for optical pumping of the electronic
population. The resulting nuclear spin polarization is both time
and position dependent. In the initial stage of the polarization
process, the induced nuclear polarization is linearly dependent on
time. For longer times the nuclear spin polarization  saturates
and is time independent. The position dependence of the induced
nuclear spin polarization is a function of the electronic
confinement across the system, and of various relaxation
mechanisms acting on the nuclear spin. Consequently, the resonance
shift induced by such a field will be position dependent.
Different experimental setups will record different resonance
shifts. For example, if the sample is grown such that in the
central region we have a $\d$-doped layer of a different nuclei
than the host nuclei, the resonance shift for such a layer will
strongly depend on its position across the well. On the other
hand, in different experiments it may be that whole magnetic
moment of the sample is recorded.

As a specific example we calculated the effects of the DNP process
for an AlGaAs PQW. The nuclear spin polarization can be as high as
80\% at $T=5$K for an initial electronic spin polarization of
100\%. The nuclear spin polarization is concentrated in the
central regions of the PWQ and depends also on temperature, being
strongly reduced as the temperature increases. The DNP effect
provides the potential to manipulate nuclear spins in
semiconductor nanostructures, making the nuclear spins an
important candidate for new electronic devices. The particular
geometry of the PQW permits a sensitive control of nuclear spins
with small electric fields.

We would like to acknowledge D. D. Awschalom and M. Poggio for
helpful discussions. Our work was supported by DARPA/ARO
DAAD19-01-1-0490.

\end{document}